\documentclass[showpacs,aps,prb,twocolumn,superscriptaddress,nofootinbib]{revtex4-1}
\usepackage{graphicx} 
\usepackage{dcolumn}
\usepackage{bm}
\usepackage{array}
\usepackage{amssymb,amsmath}
\usepackage{epstopdf}
\usepackage{xcolor}
\usepackage[normalem]{ulem}  
\usepackage[bottom]{footmisc}
\usepackage{adjustbox}
\usepackage{rotating}
\usepackage{multirow, makecell}

\newcolumntype{C}[1]{>{\centering\arraybackslash}p{#1}}

\begin{document}
\title{Comparative Raman Scattering Study of Crystal Field Excitations in Co-based Quantum Magnets}
\author{Banasree S. Mou}
\affiliation{Department of Physics and Center for Functional Materials, Wake Forest University, Winston-Salem, North Carolina 27109, USA}

\author{Xinshu Zhang}
\affiliation{Institute for Quantum Matter and Department of Physics and Astronomy, Johns Hopkins University, Baltimore, Maryland 21218, USA}

\author{Li Xiang}
\affiliation{National High Magnetic Field Laboratory, Tallahassee, Florida 32310, USA}

\author{Yuanyuan Xu}
\affiliation{Institute for Quantum Matter and Department of Physics and Astronomy, Johns Hopkins University, Baltimore, Maryland 21218, USA}

\author{Ruidan Zhong}
\affiliation{Department of Chemistry, Princeton University, Princeton, New Jersey 08544, USA}

\author{Robert J. Cava}
\affiliation{Department of Chemistry, Princeton University, Princeton, New Jersey 08544, USA}

\author{Haidong Zhou}
\affiliation{Department of Physics and Astronomy, University of Tennessee, Knoxville, Tennessee 37996, USA}

\author{Zhigang Jiang}
\affiliation{School of Physics, Georgia Institute of Technology, Atlanta, Georgia 30332, USA}

\author{Dmitry Smirnov}
\email{smirnov@magnet.fsu.edu}
\affiliation{National High Magnetic Field Laboratory, Tallahassee, Florida 32310, USA}

\author{Natalia Drichko}
\email{drichko@jhu.edu}
\affiliation{Institute for Quantum Matter and Department of Physics and Astronomy, Johns Hopkins University, Baltimore, Maryland 21218, USA}

\author{Stephen M. Winter}
\email{winters@wfu.edu}
\affiliation{Department of Physics and Center for Functional Materials, Wake Forest University, Winston-Salem, North Carolina 27109, USA}
\date{\today}

\newcommand\sw[1]{\textcolor{blue}{(SW: #1)}}
\newcommand\note[1]{\textcolor{blue}{(note: #1)}}

\newcommand\DS{\textcolor{magenta}}

\begin{abstract}
Co-based materials have recently been explored due to potential to realise complex bond-dependent anisotropic magnetism. Prominent examples include Na$_2$Co$_2$TeO$_6$, BaCo$_2$(AsO$_4$)$_2$, Na$_2$BaCo(PO$_4$)$_2$, and CoX$_2$ (X = Cl, Br, I). In order to provide insight into the magnetic interactions in these compounds, we make a comparative analysis of their local crystal electric field excitations spectra via Raman scattering measurements. Combining these measurements with theoretical analysis confirms the validity of $j_{\rm eff} = 1/2$ single-ion ground states for all compounds, and provides accurate experimental estimates of the local crystal distortions, which play a prominent role in the magnetic couplings between spin-orbital coupled Co moments. 
\end{abstract}

\maketitle

\section{Introduction}

The Kitaev spin liquid model has attracted the attention of the condensed matter community already for more than a decade\cite{Kitaev2006,Takagi2019}. This model of $S=1/2$ spins on a honeycomb  lattice stands apart as exactly solvable, which has allowed a great depth of understanding of the ground state and excitations. Shortly after solution of the model, it was proposed that magnetic atoms with strong spin-orbit coupling (SOC) can realise dominant Kitaev interactions\cite{jackeli2009mott}, with Ir$^{4+}$-  and Ru$^{3+}$-based compounds as the first actively studied candidates\cite{singh2010antiferromagnetic,singh2012relevance,kim2015kitaev,sears2015magnetic}.  Recently, this interest extended to high spin Co$^{2+}$ $d^7$ compounds, due to their potential to realise bond-dependent anisotropic exchange interactions \cite{kim2021spin,doi:10.1142/S0217979221300061,watanabe2022frustrated}. In these materials, SOC results in a $j_{\rm eff} = 1/2$ single-ion ground state, for which a significant orbital moment affords large anisotropies in the magnetic couplings\cite{Sano2018PRB,Liu2018,Liu2020PRL,liu2023non}.  Such bond-dependent couplings can enhance quantum fluctuations, in cases where they compete on different bonds, such as extended versions of Kitaev's honeycomb model \cite{Winter2017a,Hermanns2017,Takagi2019,Trebst2022}, or other quantum compass models\cite{nussinov2015compass}. 

\begin{figure}[t]
\includegraphics[width=0.8\linewidth]{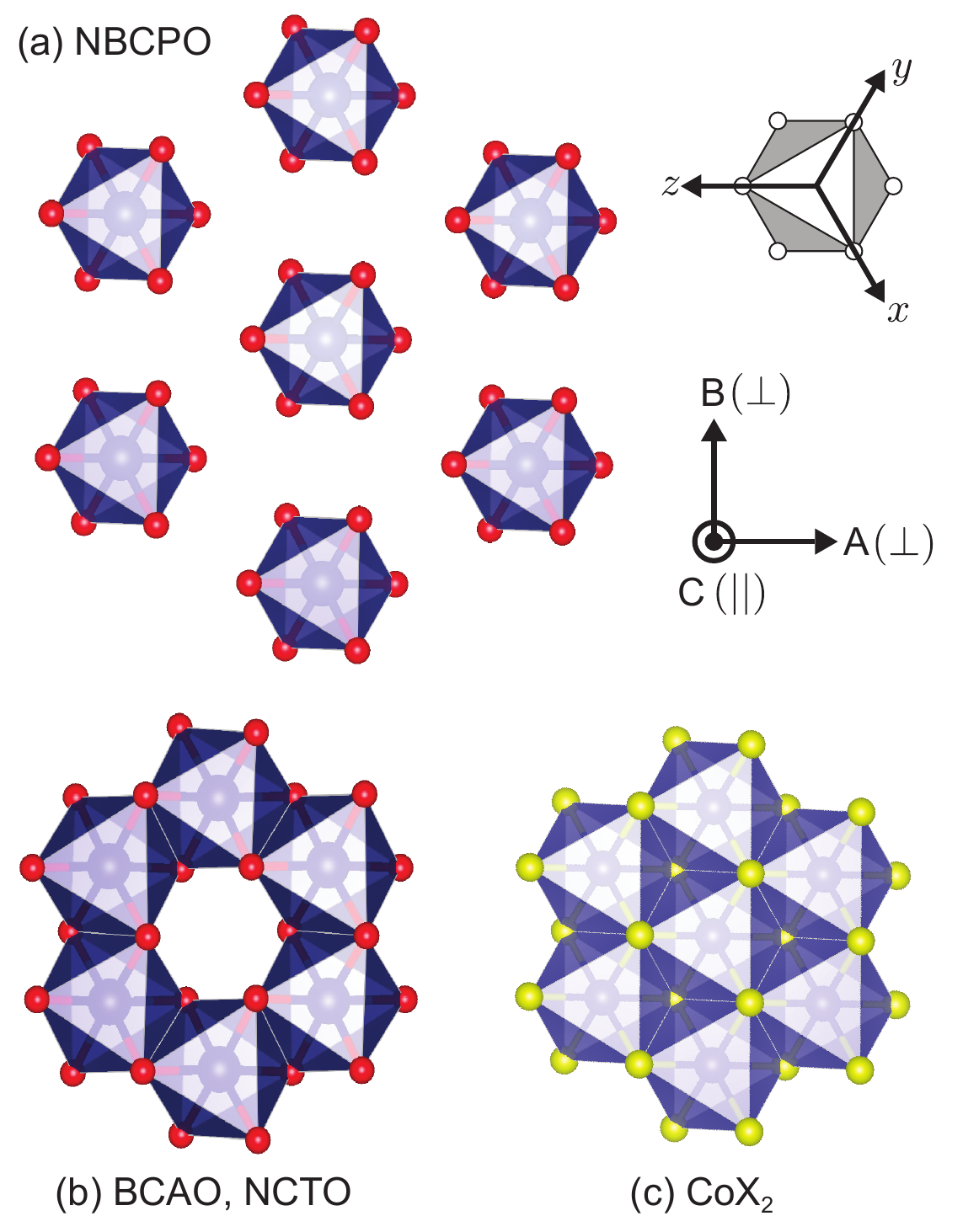}
\caption{Co$^{2+}$ octahedra in studied compounds showing global cubic ($xyz$) and ABC coordinates. (a) Isolated triangular lattice in Na$_2$BaCo(PO$_4$)$_2$ (NBCPO). (b) Edge-sharing honeycomb lattice in BaCo$_2$(AsO$_4$)$_2$ (BCAO) and Na$_2$Co$_2$TeO$_6$ (NCTO). (c) Edge-sharing triangular lattice in CoX$_2$ (X = Cl, Br, I). }
\label{fig-structures}
\end{figure}

Interest into such effects has led to exploration of the magnetic properties of various materials, such as the zigzag chain compound CoNb$_2$O$_6$\cite{fava2020glide,morris2021duality,woodland2023tuning}, the honeycomb lattice compounds BaCo$_2$(AsO$_4$)$_2$ (BCAO)\cite{zhong2020weak,zhang2023magnetic} and Na$_2$Co$_2$TeO$_6$ (NCTO)\cite{Lefrancois2016,Kim2021,Yao2022}, and the triangular lattice compounds Na$_2$BaCo(PO$_4$)$_2$ (NBCPO)\cite{zhong2019strong,li2020possible,wellm2021frustration} and CoI$_2$\cite{kim2023bond}. The latter four, which are the subject of this work, are depicted in Fig.~\ref{fig-structures}. For several of these compounds, there has been significant recent discussion over the precise low-energy magnetic models \cite{Songvilay2020PRB,Sanders2022PRB,halloran2023geometrical}, which may be hard to constrain without input from a variety of experiments. 

A crucial controlling factor in all Co $3d^7$ compounds is the local crystal field splitting, which results from distortion of the octahedra, and competes with SOC to determine the specific composition of the local moments\cite{kim2021spin,winter2022magnetic,liu2023non}. In contrast to heavier $4d$ and $5d$ compounds such as ruthenates and iridates, SOC is sufficiently weak in Co materials that even small departures from ideal octahedral crystal fields can lead to significant modification of the magnetic couplings. For this reason, accurate understanding of the crystal field splitting is important for derivation of underlying microscopic models. In this work, we utilize Raman scattering measurements to probe the low-lying crystal-field excitations between different spin-orbital states in the prominently  studied Co oxides depicted in Fig.~\ref{fig-structures}. Analysis aided by model calculations of the Raman scattering intensities and energies allows us to provide precise estimates of the trigonal crystal field strength and anisotropic $g$-tensors. These results provide direct rationalization of effects such as (a) the weak Ising anisotropy in NBCPO, (b) the strong bond-independent easy-plane anisotropy in BCAO, and (c) the near absence of XXZ anisotropy in CoI$_2$. 

The paper is organized as follows. In Section \ref{sec_methods}, we provide the experimental details of the Raman scattering measurements, as well as the modelling of the local crystal-field excitations and experimental fitting. In Section \ref{sec_model} we discuss the theoretical evolution of the electronic Raman spectrum with the trigonal crystal field. In Section \ref{sec_analysis}, we present the Raman spectra and fitted parameters for each compound. Finally, the results are summarized in Section \ref{sec_conclusions}.

\section{Methods} \label{sec_methods}

\subsection{Experimental Details}

Raman scattering measurements of Na$_2$BaCo(PO$_4$)$_2$ and  BaCo$_2$(AsO$_4$)$_2$  single crystals were performed using the Jobin-Yvon T64000 triple monochromator spectrometer equipped with a liquid nitrogen cooled CCD detector. Spectral resolution was 2 cm$^{-1}$. 514.5 nm and 647 nm  lines of Ar$^+$-Kr$^+$ mixed gas laser were used as the excitation light. The measurements were performed in  pseudo-Brewster's  geometry using laser probe with the power of 10 mW  focused into an elliptically shaped spot of 50 by 100 $\mu$m in size. The polarization-resolved spectra were measured from the $ab$  planes of the crystals in two configurations: $(xx)$ and  $(xy)$. Polarizations were determined by orienting single crystals of the measured samples in the  beam in such a way that  the intensity of A$_{1g}$ phonons in $(xy)$ scattering channel was minimized.  For low temperature measurements the sample was  mounted on the cold-finger of Janis ST-500 cryostat.  The presented Raman response $\chi''(\omega, T)$ was normalized on the Bose-Einstein factor $[n(\omega, T)+1]$, where $n(\omega, T) = [\exp(\hbar \omega / k_\mathrm{B}T) - 1]^{-1}$ is the Bose occupation factor.

Raman spectroscopic measurements were performed on single-crystal Na$_2$Co$_2$TeO$_6$ where the temperature control and applied magnetic field up to 14T are done via a Quantum Design physical property measurement system (PPMS). The scattered light was collected in backscattering geometry from the $ab$ plane of a Na$_2$Co$_2$TeO$_6$ crystal with the magnetic field applied along the $c$ axis (Faraday geometry) using an unpolarized 532 nm laser focused to a (1-2)-$\mu$m-diameter spot with the incident power of $\sim$1 mW. The collected signal spectra were analyzed by a monochromator (SP2750 Princeton Instruments, 1100 g/mm grating) and recorded by a liquid-nitrogen cooled CCD (PyLoN100BR, Princeton Instruments) with a spectral resolution of about 0.7 cm$^{-1}$. The same setup was used to perform measurements on the single crystals of BaCo$_2$(AsO$_4$)$_2$  at $T=2$ K and magnetic fields $H\parallel c$ up to 14 T.

\subsection{Computational Details}

\begin{figure}[t]
\includegraphics[width=0.9\linewidth]{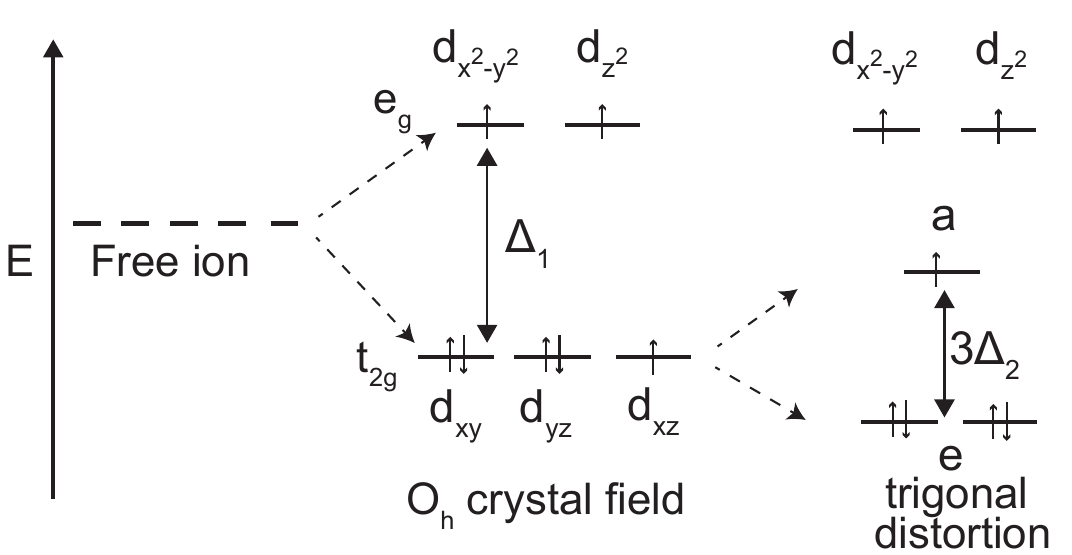 }
\caption{Splitting of single-particle orbitals with octahedral and trigonal crystal fields (SOC not included). }
\label{fig-energy}
\end{figure}

To obtain theoretical Raman intensities for the local crystal-field transitions, we exactly diagonalized a simplified one-site model including $\mathcal{H} = \mathcal{H}_{\rm CFS} + \mathcal{H}_{\rm SOC} + \mathcal{H}_U$, which is the sum of crystal field, spin-orbit coupling, and Coulomb terms, respectively. For the crystal-field Hamiltonian, we consider an ideal trigonal distortion retaining $D_{3d}$ site symmetry. The Hamiltonian can be written:
\begin{align}
\mathcal{H}_{\rm CFS} = \sum_{\sigma} \mathbf{c}_{i,\sigma}^\dagger \ \mathbb{D} \ \mathbf{c}_{i,\sigma}
\end{align}
where:
\begin{align}
\mathbf{c}_{i,\sigma}^\dagger = \left(c_{i,yz,\sigma}^\dagger \  c_{i,xz,\sigma}^\dagger \ c_{i,xy,\sigma}^\dagger \ c_{i,z^2,\sigma}^\dagger \ c_{i,x^2-y^2,\sigma}^\dagger\right)
\end{align}
creates an electron in one of the $d$-orbital (Wannier) functions at site $i$. 
In terms of the global cubic $(xyz)$ coordinates defined in Fig.~\ref{fig-structures}, the CFS matrix is given by:
\begin{align}
\mathbb{D} = \left( \begin{array}{ccccc} 
0 & \Delta_2 & \Delta_2 & 0 & 0 \\
\Delta_2 & 0 & \Delta_2 & 0 & 0 \\
\Delta_2 & \Delta_2 & 0 & 0 & 0 \\
0 & 0 & 0 & \Delta_1 & 0 \\
0 & 0 & 0 & 0 & \Delta_1
\end{array}\right)
\end{align}
where $\Delta_1$ is the $t_{2g}$-$e_g$ splitting, and $\Delta_2$ is the trigonal term. The effect of these crystal field terms on the orbital energies is depicted in Fig.~\ref{fig-energy}. The SOC term is written:
\begin{align}
\mathcal{H}_{\rm SOC} = \lambda \mathbf{L}_i \cdot \mathbf{S}_i
\end{align}
where the atomic SOC constant for Co is roughly $\lambda_{\rm Co} \approx 60$ meV.  In the analysis in section \ref{sec_analysis}, we let $\lambda$ be a fitting parameter, but find only small deviations from the expected atomic value. The Coulomb interactions are most generally written:
\begin{align}
\mathcal{H}_U = \sum_{\alpha,\beta,\delta,\gamma}\sum_{\sigma,\sigma^\prime}U_{\alpha\beta\gamma\delta} \ c_{i,\alpha,\sigma}^\dagger c_{i,\beta,\sigma^\prime}^\dagger c_{i,\gamma,\sigma^\prime} c_{i,\delta,\sigma}
\end{align}
where $\alpha,\beta,\gamma,\delta$ label different $d$-orbitals. 
In the spherically symmetric approximation \cite{sugano2012multiplets}, the coefficients $U_{\alpha\beta\gamma\delta}$ are all related to the three Slater parameters $F_0, F_2, F_4$. In terms of these, the familiar $t_{2g}$ Kanamori parameters are $
U_{t2g} = F_0 + \frac{4}{49} \left( F_2 + F_4 \right)$ and 
$J_{t2g} = \frac{3}{49} F_2 + \frac{20}{441} F_4$.
Throughout, we use $U_{t2g} =4.25$ eV, $J_{t2g} = U_{t2g}/5$, and $F_4/F_2 = 5/8$, which is appropriate for $3d$ elements \cite{pavarini2014dmft}.

The Raman scattering intensity at frequency $\omega$ is generally given by:
\begin{align}
    I(\omega) \propto & \ \left|\langle n| \mathcal{R}(\omega_0) |g\rangle \right|^2\delta(\omega-E_n+E_g) 
    \\
    \mathcal{R}(\omega_0) = & \  \sum_m\frac{(\hat{e}_{\rm out} \cdot \vec{\mu} ) |m\rangle \langle m| (\hat{e}_{\rm in} \cdot \vec{\mu})}{\omega_0-E_m+E_g}
\end{align}
where $\omega_0$ is the Raman laser frequency, $|g\rangle$ is the ground state, $|n\rangle$ is the excited state, and $\hat{e}_{\rm in}$ and $\hat{e}_{\rm out}$ are the incident and scattered polarizations. Here, $\vec{\mu}$ is the electric dipole operator. In the standard Uns\"old approximation \cite{unsold1927quantentheorie,boyd2020nonlinear}, it is assumed that $\omega_0$ is not nearly resonant with any particular state $|m\rangle$, and $E_m - E_g \gg \omega_0$, such that the summation can be approximated:
\begin{align}
 \mathcal{R}(\omega_0) \approx & \ (\hat{e}_{\rm out} \cdot \vec{\mu} )  (\hat{e}_{\rm in} \cdot \vec{\mu})
\end{align}
with electric dipole operator given by:
\begin{align}
\vec{\mu} = e \sum_{\alpha,\beta,\sigma}\langle d_{i,\alpha}| \vec{r}| d_{i,\beta}\rangle c_{i,\alpha,\sigma}^\dagger c_{i,\beta,\sigma}
\end{align}
We take the position matrix elements between $d$-orbital Wannier functions to be equal to those between equivalent hydrogenic orbitals.

In order to estimate the crystal field terms from {\it ab-initio}, we performed fully relativistic density functional theory (DFT) calculations at the GGA (PBE \cite{perdew1996generalized}) level using FPLO \cite{koepernik1999full,opahle1999full}. For this purpose, we employ a 12$\times$12$\times$12 grid, and crystal field and hopping parameters were extracted via Wannier fitting. For each material, the starting geometry was taken from published crystal structures: Na$_2$BaCo(PO$_4$)$_2$ \cite{zhong2019strong}, BaCo$_2$(AsO$_4$)$_2$ \cite{djordevic2008baco2}, CoBr$_2$ and CoCl$_2$ \cite{wilkinson1959neutron}, and CoI$_2$ \cite{wyckoff1963crystal}. The only exception is for NCTO, for which we modelled the random positions of the statistically disordered Na atoms by performing calculations on five different structures representing the breadth of possible local geometries (see Ref.~\onlinecite{xiang2023disorder} for further details.) Computed values of $\Delta_2$ are shown in Table \ref{tab:data}. As we discuss below, the DFT results reproduce trends in $\Delta_2$ within families of compounds, but do not reproduce experimental values for the halide materials within required accuracy for modelling magnetic response. As a consequence, experimental input is of significant value.

For the purpose of fitting experimental data, we exactly diagonalized the one-site model for $\Delta_1 = 1.0$ eV, and a range of $\Delta_2$ and $\lambda$ values. The low-lying spin-orbit excitons are not strongly influenced by $\Delta_1$, so this value is suitable for analysis of both oxides and halides. The numerical excitation energy data was then employed to construct analytical interpolation functions, which were fit to the experimental peak energies. 

\section{Theoretical Evolution of Electronic Raman Scattering}\label{sec_model}

\begin{figure}[t]
\includegraphics[width=\linewidth]{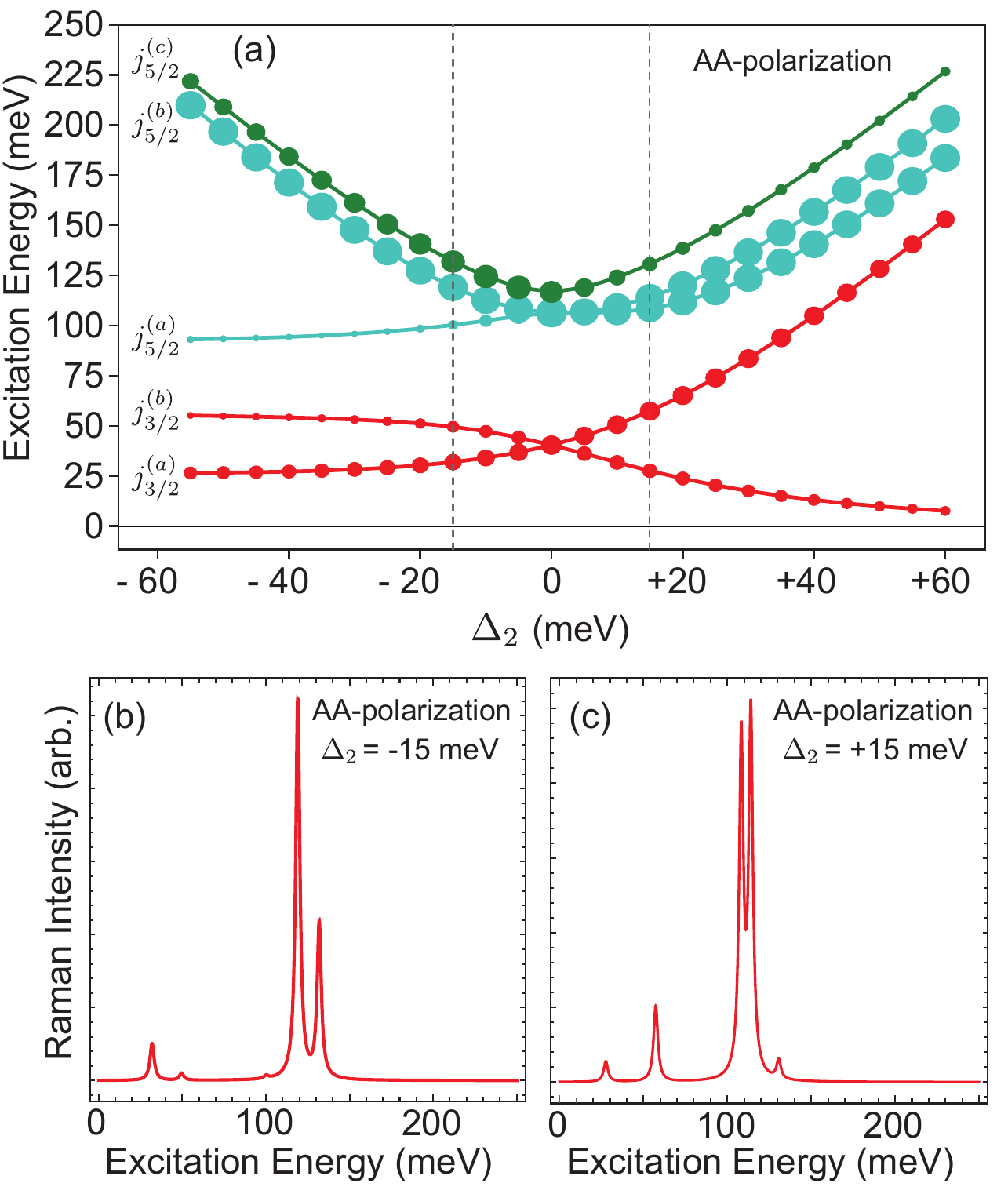 }
\caption{(a) Theoretical evolution of the electronic Raman scattering intensity for $\lambda = 60$ meV, as a function of trigonal term $\Delta_2$. Relative intensity of each peak is proportional to the area of the circles. Results are shown for the AA-polarization; the AB, and BB polarizations are essentially identical. Coordinates are defined in Fig.~\ref{fig-structures}. (b,c) Simulated spectra for $\Delta_2 = \pm 15$ meV to show expected intensity distributions. For large $|\Delta_2|$, some peaks are sufficiently weak as to be difficult to resolve experimentally.}
\label{fig-theoryraman}
\end{figure}

In order to interpret the experimental spectra, we first investigate theoretically the low-energy electronic Raman scattering intensity as a function of $\Delta_2$, as shown in Fig.~\ref{fig-theoryraman}. For the purpose of discussion, we take $\lambda_{\rm Co} = 60$ meV, and $\Delta_1 = $ 1.0 eV, which is a realistic range.

It is useful to first consider the effects of the trigonal distortion in the absence of SOC. The resulting splitting of the $d$-orbitals with $\Delta_1$ and $\Delta_2$ is shown in Fig.~\ref{fig-energy}. The primary effect of $\Delta_2$ is to split the $t_{2g}$ orbitals into $e$ and $a$ combinations. In the limit of large positive $\Delta_2\gg 0$, the three unpaired electrons of the high-spin $d^7$ configuration would nominally reside in the $d_{x^2-y^2}$ and $d_{z^2}$ orbitals and the $a$ combination of the $t_{2g}$ orbitals, as pictured in Fig.~\ref{fig-energy}. Only in this limit is the orbital angular momentum completely quenched. Otherwise, for small $|\Delta_2| \ll \lambda$ or $\Delta_2<0$, there remains an unquenched orbital degree of freedom. In such case, the overall single-ion ground state is a spin-orbital multiplet, whose precise composition varies with $\Delta_2/\lambda$. For all values of $\Delta_2$, the ground state is a doublet. The evolution of the excitations energies to higher lying states as a function of $\Delta_2$ is shown in Fig.~\ref{fig-theoryraman}(a).

In the absence of trigonal splitting (i.e.~for $\Delta_2 = 0$), the spin-orbital multiplets are nominally split into a $j_{1/2}$ ground state, and $j_{3/2}$ and $j_{5/2}$ excited states. Within the $O_h$ double group, the $j_{1/2}$ states transform as the doublet $\Gamma_6$ representation, the $j_{3/2}$ states transform as the quartet $\Gamma_8$ representation, and the $j_{5/2}$ states transform as $\Gamma_7\oplus \Gamma_8$. Note that the $j_{5/2}$ states are already slightly split even in a perfect $O_h$ symmetry. All electronic transitions from the $j_{1/2}$ to $j_{3/2}$ ($\Gamma_6 \to \Gamma_8$) and from $j_{1/2}$ to $j_{5/2}$ ($\Gamma_6 \to \Gamma_7$ and $\Gamma_8$) are Raman active, and thus have finite intensity. These intensities were found to depend only weakly on in-plane polarization. 

For finite $\Delta_2$, the symmetry is reduced to $D_{3d}$ or lower. All quartets are split into doublets. In this case, it is convenient to still refer to the nominal symmetries of the transitions within the $O_h$, even though such symmetries do not apply strictly. The particular pattern of intensities then depends on the sign of $\Delta_2$, which makes Raman scattering a particularly useful diagnostic probe of Co(II) materials. For example, for $\Delta_2 > 0$, the two $j_{1/2}$ to $j_{5/2}$ transitions of lower energy are anticipated to have far greater intensity than the highest energy transition. As a consequence, even for relatively small distortions, the sign of $\Delta_2$ can, in principle, be determined spectroscopically. In practice (as discussed in the next section), it may not always be possible to resolve all excitations, in which case the assignment of the observed transitions can still be made by referring to the modes expected to have the highest intensities. 

\section{Analysis of Materials (experimental results)}
\label{sec_analysis}

\begin{table}
\caption{\label{tab:data}Summary of experimental crystal field excitations, fitted values of trigonal splitting $\Delta_2$ and SOC constant $\lambda$. DFT estimates of $\Delta_2$ were obtained via projective Wannier fitting as described in the text. All energies are in units of meV. The symmetry labels $\Gamma_n$ refer to the $O_h$ double group, and apply strictly only in the absence of trigonal distortion. }

\begin{ruledtabular}

\begin{tabular}{| l | c |c | c |}
     & \multirow{2}{*}{Excitation}      & Exp.  & Fitted\\ 
     &                                  & Energies      & Energies \\

    \hline
     \textbf{ Na$_2$BaCo(PO$_4$)$_2$}  &$j_{3/2}^{(a)}$ ($\Gamma_8$) & 39 & 39 \\ 
       { $\Delta_2$(DFT)} = -4.4 meV & $j_{3/2}^{(b)}$ ($\Gamma_8$)  & 44 & 45\\
       { $\Delta_2$(fit)} = -3.7 meV & $j_{5/2}^{(a)}$ ($\Gamma_8$)  & $-$ & 110 \\
       { $\lambda$(fit)}  = 62.8 meV & $j_{5/2}^{(b)}$ ($\Gamma_8$)  & 111 & 112 \\
                             & $j_{5/2}^{(c)}$ ($\Gamma_7$)  & 125 & 123  \\

      \hline
     \textbf{ BaCo$_2$(AsO$_4$)$_2$}  & $j_{3/2}^{(a)}$ ($\Gamma_8$) & $\sim17.5$ & 15 \\ 
        { $\Delta_2$(DFT)} = + 43.9 meV& $j_{3/2}^{(b)}$ ($\Gamma_8$)  &  $-$ & 105\\
        { $\Delta_2$(fit)} = + 40 meV & $j_{5/2}^{(a)}$ ($\Gamma_8$) & 143 & 143 \\
        { $\lambda$(constrained)}  = 63 meV & $j_{5/2}^{(b)}$ ($\Gamma_8$)  & 156 & 159\\
     & $j_{5/2}^{(c)}$ ($\Gamma_7$) & $-$ &  182 \\
     
     \hline
      \textbf{ Na$_2$Co$_2$TeO$_6$}\cite{chen2021spin}  &$j_{3/2}^{(a)}$ ($\Gamma_8$) & 22 & 24 \\ 
        { $\Delta_2$(DFT)} = [2.4 meV & $j_{3/2}^{(b)}$ ($\Gamma_8$) & 69 & 74 \\
        to 18.5 meV]                  & $j_{5/2}^{(a)}$ ($\Gamma_8$) & 118  & 122  \\
        { $\Delta_2$(fit)} = +23.7 meV & $j_{5/2}^{(b)}$ ($\Gamma_8$)  & 138  & 132 \\
         { $\lambda$(fit)}  = 64.4 meV&$j_{5/2}^{(c)}$ ($\Gamma_7$) & $-$ & 152  \\

     \hline

      \textbf{ CoCl$_2$} \cite{christie1971electronic,christie1975electronic} & $j_{3/2}^{(a)}$ ($\Gamma_8$) & 29 & 29 \\ 
        { $\Delta_2$(DFT)} = -5.8 meV& $j_{3/2}^{(b)}$ ($\Gamma_8$)  & 68 & 64\\
        { $\Delta_2$(fit)} = + 18.3 meV& $j_{5/2}^{(a)}$ ($\Gamma_8$) & 119 & 116 \\
        { $\lambda$(fit)}  = 63.6 meV& $j_{5/2}^{(b)}$ ($\Gamma_8$)  & 122 & 123 \\
     & $j_{5/2}^{(c)}$ ($\Gamma_7$) & 138 &  142 \\
      
     \hline
     \textbf{ CoBr$_2$} \cite{lockwood1979raman} & $j_{3/2}^{(a)}$ ($\Gamma_8$) & 33 & 34  \\ 
        { $\Delta_2$(DFT)} = -9.4 meV& $j_{3/2}^{(b)}$ ($\Gamma_8$)  & 54 & 53\\
        { $\Delta_2$(fit)} = + 9.9 meV& $j_{5/2}^{(a)}$ ($\Gamma_8$) & 115 & 113 \\
        { $\lambda$(fit)}  = 63.6 meV& $j_{5/2}^{(b)}$ ($\Gamma_8$)  & 117 & 115 \\
     & $j_{5/2}^{(c)}$ ($\Gamma_7$) & 127 &  131 \\
      
     \hline
     \textbf{ CoI$_2$} \cite{mischler1987raman} & $j_{3/2}^{(a)}$ ($\Gamma_8$) & 27 & 35 \\ 
        { $\Delta_2$(DFT)} = -21 meV& $j_{3/2}^{(b)}$ ($\Gamma_8$)  & 41 & 47\\
        { $\Delta_2$(fit)} = + 6.7 meV & $j_{5/2}^{(a)}$ ($\Gamma_8$) & 109 & 106 \\
        { $\lambda$(fit)}  = 60 meV& $j_{5/2}^{(b)}$ ($\Gamma_8$) & 111 & 107 \\
     &$j_{5/2}^{(c)}$ ($\Gamma_7$) & 118 & 120  \\

\end{tabular}
\end{ruledtabular}
\end{table}

In this section, we present and analyze experimental electronic Raman scattering spectra for a number of Co(II) compounds. A summary of excitation energies and relative intensities is given in Table \ref{tab:data}. 

\begin{figure}[t]
\includegraphics[width=0.80\linewidth]{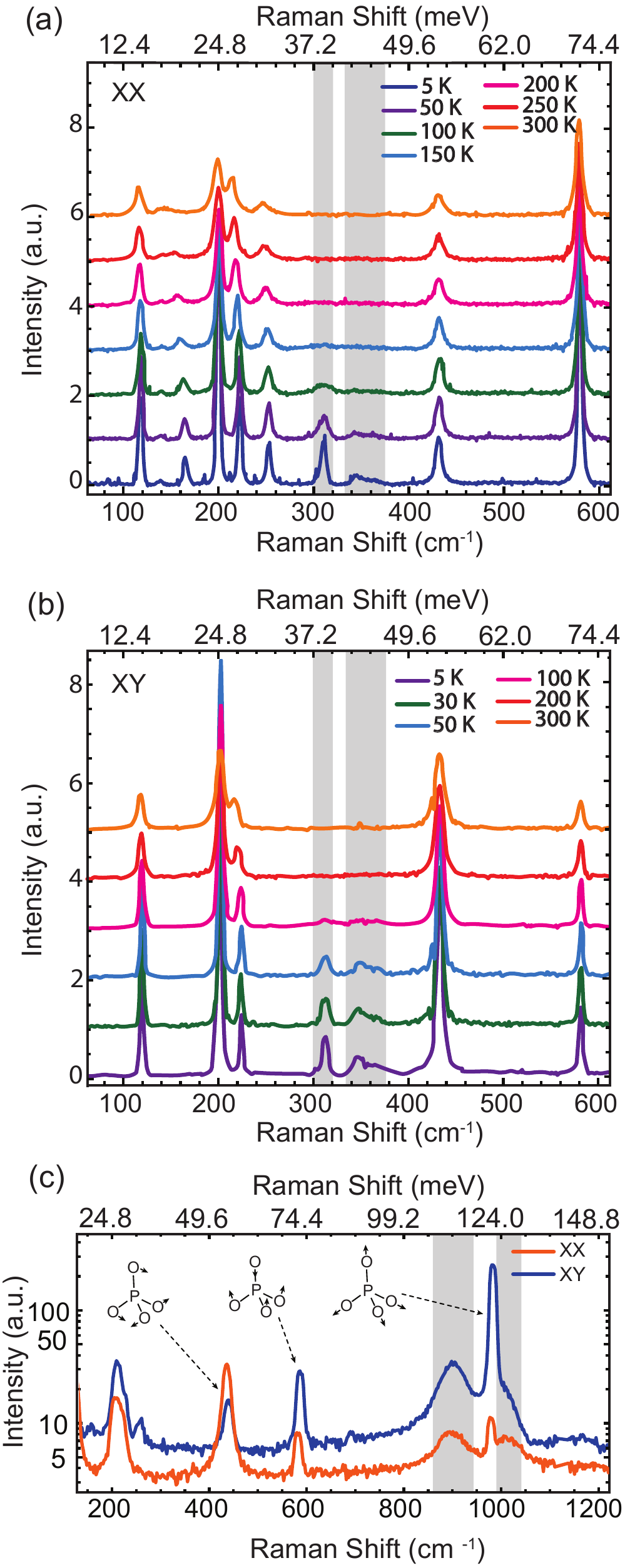}
\caption{Experimental Raman intensities for NBCPO. The shaded regions are CF excitations. (a, b) Temperature-dependence of spectra in the range of 20 - 600 cm$^{-1}$ in xx and xy polarization, respectively. (c) Room temperature spectra in range of 150 - 1200 cm$^{-1}$, with local [PO$_4$]$^{3-}$ vibrations depicted.}
\label{fig-nbcpo}
\end{figure}

{\it Na$_2$BaCo(PO$_4$)$_2$}: In NBCPO, the Co$^{2+}$ form a triangular lattice of isolated octahedra \cite{zhong2019strong}. The material exhibits a complex phase diagram as a function of temperature and field \cite{lee2021temporal,sheng2022two} including strong magnetocaloric effects \cite{liu2022quantum}. Initial thermal conductivity measurements \cite{li2020possible} were interpreted in terms of gapless excitations, but more recent data \cite{huang2022thermal} challenged this interpretation. 

The relative geometry of neighboring octahedra in NBCPO is the same as for third neighbors in the edge-sharing honeycomb lattice compounds like BCAO, as shown in Fig.~\ref{fig-structures}. The magnetic couplings for such a geometry have been previously analyzed\cite{winter2022magnetic}, and are anticipated to be dominated by a single exchange process between electrons occupying the $e_g$ orbitals, which leads to bond-independent XXZ couplings of the form $\mathcal{H}_{ij} = J_{\perp}(S_i^x S_j^x + S_i^y S_j^y) + J_{||} S_i^z S_j^z$. Here, $x$ and $y$ refer to in-plane directions, and $z$ refers to the out-of-plane direction. The anisotropy in both the magnetic couplings $J$ and gyromagnetic factors $g$ originate from ``squeezing'' of the local $j_{1/2}$ moments via the trigonal crystal-field $\Delta_2$. It can be anticipated that $\Delta_2 >0$ corresponds to XY anisotropy $J_{||}/J_{\perp} < 1$ (and $g_{||}<g_\perp$), while $\Delta_2 < 0$ corresponds to Ising anisotropy $J_{||} /J_{\perp} > 1$ (and $ g_{||} > g_\perp$). Assessment of $\Delta_2$ is therefore relevant to the magnetic model.

Experimental Raman scattering spectra of Na$_2$BaCo(PO$_4$)$_2$ for polarization in the $ab$ plane are presented in  Fig.~\ref{fig-nbcpo}. The temperature dependence of the lower energy modes is shown in Fig.~\ref{fig-nbcpo}(a,b). Sharp features of $\Gamma$-point phonons  are observed in the whole measured temperature range, while the electronic crystal-field excitations at 39 and 44 meV are observed as distinct bands in the spectra only below 50~K. These can be identified by the strong temperature dependence of their intensity, and,  in part, from the weak polarization-dependence. Fig.~\ref{fig-nbcpo}(c) shows room temperature spectra covering the higher energy range. Sharp phonons in this region can be identified as local [PO$_4$]$^{3-}$ bending (440 cm$^{-1}$) and stretching (590, 980 cm$^{-1}$) modes (see, for example, Ref.~\onlinecite{inorg2009,lee2008raman,zhai2011high}). Broad electronic modes are seen at 111, and 125 meV. 

We assign the crystal-field transitions as follows: The narrowly split lower energy pair at 39 and 44 meV are assigned as $\Gamma_6 \to \Gamma_8$ transitions (i.e.~$j_{1/2}\to j_{3/2}$). On the basis of their relative intensities, we interpret that $\Delta_2 <0$. In this case, we would expect the $j_{1/2} \to j_{5/2}$ transitions to have dominant intensity in the two higher energy modes, nominally with $\Gamma_6 \to \Gamma_8$ and $\Gamma_6 \to \Gamma_7$ symmetries. We assign the peaks at 111 and 125 meV to these transitions. In principle, another $\Gamma_6 \to \Gamma_8$ transition with very weak intensity would be expected to lie just below the 111 meV mode [see Fig.~\ref{fig-theoryraman}(b)], but this likely overlaps with the more intense 111 meV excitation and is therefore not separately distinguishable. Fitting the excitation energies yields the experimental estimates of $\Delta_2\text{(fit)} = -3.7$ meV and $\lambda\text{(fit)} = 63$ meV, which are remarkably consistent with the {\it ab-initio} values shown in Table \ref{tab:data}. This constitutes a very weak trigonal term, such that NBCPO should be viewed as a nearly ideal $j_{\rm eff} = 1/2$ system. On the basis of the fitted parameters, we compute $g_{\perp} = 4.26$ and $g_{||} = 4.83$, in excellent agreement with the experimental values of 4.21 and 4.81, respectively, from electron spin resonance (ESR) \cite{wellm2021frustration}. It may therefore be anticipated that NBCPO exhibits a {\it weak} Ising anisotropy, which is consistent with inelastic neutron scattering data \cite{sheng2022two}, which provided an estimate of $J_{||} / J_{\perp} = 1.73$, and showed the emergence of gapped magnons under field. Although this degree of anisotropy may seem large for such a weak crystal field splitting, the ratio $J_{||} / J_{\perp}$ is theoretically unbounded \cite{winter2022magnetic}, and much stronger Ising anisotropies are possible for more significant distortions. The finding that even relatively small trigonal crystal fields with $|\Delta_2/\lambda| \sim 0.06$ can have significant influence on the magnetic anisotropy highlights the importance of understanding the local crystal field effects for interpreting the magnetic response of Co$^{2+}$ compounds.

{\it BaCo$_2$(AsO$_4$)$_2$}: In BCAO, the CoO$_6$ octahedra form an edge-sharing honeycomb lattice \cite{djordevic2008baco2}. The magnetic ground state has been the subject of some discussion, with evidence for closely related spiral and double-stripe patterns \cite{regnault1977magnetic,regnault2006investigation,regnault2018polarized}. Under field, it exhibits a series of phase transitions \cite{zhong2020weak,shi2021magnetic,tu2022evidence}, which are strongly pressure-tunable \cite{huyan2022hydrostatic}, highlighting the sensitivity of the magnetic Hamiltonian to structural details. The ESR response shows unusually broad features under applied field \cite{zhang2023magnetic}. Initial neutron scattering studies \cite{regnault1977magnetic,regnault2006investigation,regnault2018polarized} indicated significant XY-anisotropy (i.e.~$J_{\perp}/J_{||} > 1$), which is compatible with theoretical studies \cite{das2021xy,winter2022magnetic,maksimov2022ab}. A subsequent comprehensive neutron scattering study concluded that a $J_1-J_3$ XXZ exchange model with $J_{||}/J_\perp = 0.16$ provided the best explanation of the phase diagram and excitations \cite{halloran2023geometrical}. In this model, the bond-independent XXZ anisotropy was found to be significantly stronger than bond-dependent anisotropic couplings like the celebrated Kitaev coupling. 
However, this model has been called into question recently, as it was suggested that quantum fluctuations drive the model into the adjacent ferromagnetic phase instead \cite{jiang2023quantum}. In order to constrain the magnetic model, it is therefore important to establish the microscopic parameters from experiment.

\begin{figure}[t]
\includegraphics[width=0.95\linewidth]{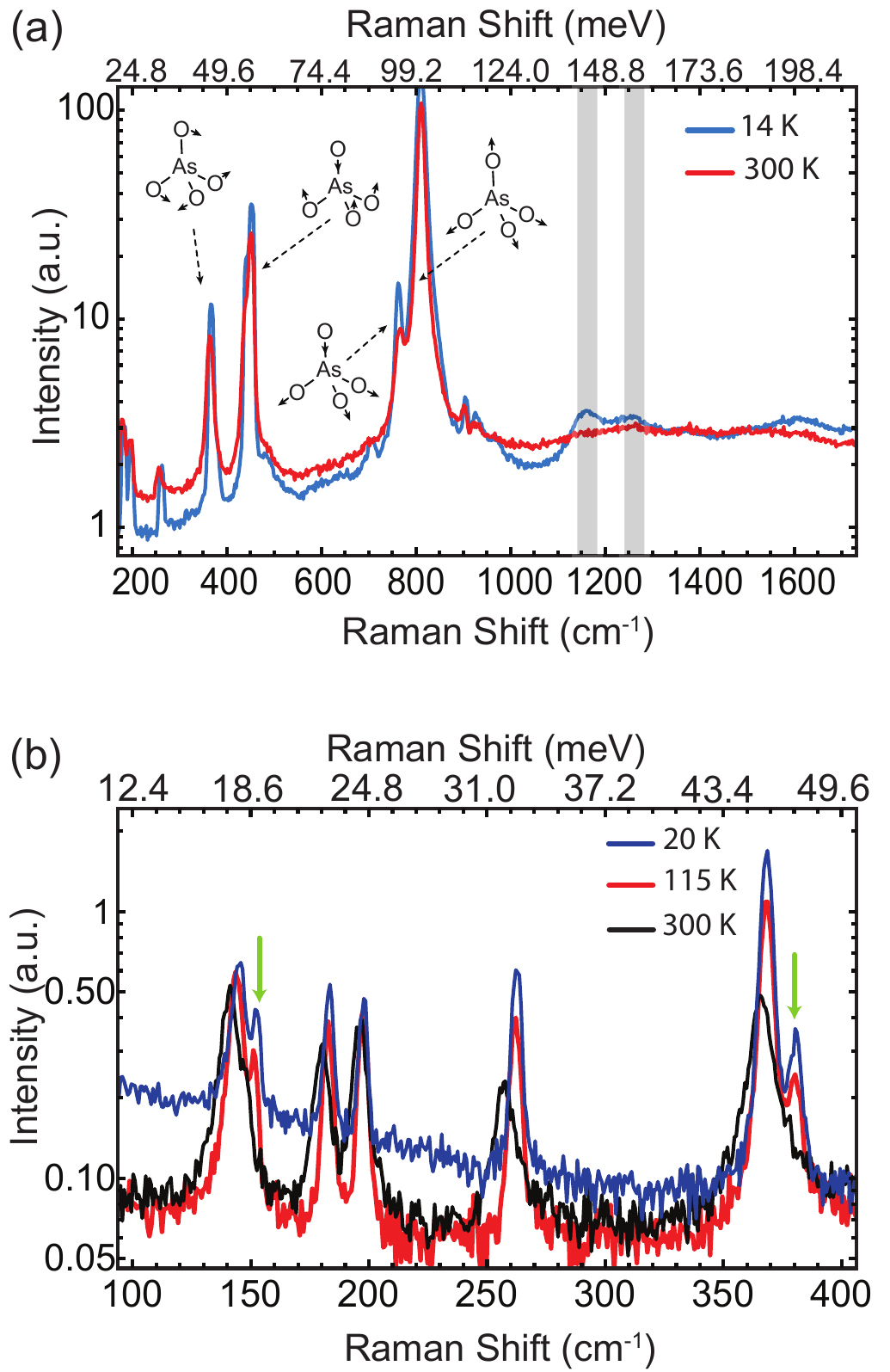}
\caption{Experimental temperature dependent Raman intensities for BCAO in  xx polarization configuration. The shaded regions are CF excitations. (a) Contrasting high-energy spectra at 14 and 300 K. Local [AsO$_4$]$^{3-}$ vibrational modes are indicated. (b) Low energy Raman spectra. Arrows indicate strongly temperature-dependent modes, which we identify as phonons (see text).}
\label{fig-bcao}
\end{figure}

In order to identify possible crystal field excitations, multiple Raman scattering experiments were performed on BCAO at a variety of temperatures and magnetic fields. In Fig.~\ref{fig-bcao}(a), we first show higher frequency zero-field Raman spectra measured with the resolution of 8 cm$^{-1}$ at temperatures 300 and 14 K. Previous analysis of the phonon modes can be found in Ref.~\onlinecite{zhang2023magnetic}. The high frequency region is populated by local vibrations which can be identified as [AsO$_4$]$^{3-}$ bending (370, 450 cm$^{-1}$) and stretching modes (780, 810 cm$^{-1}$), in accordance with Ref.~\onlinecite{inorg2009,frost2010raman}. The broad features at 1154 and 1260 cm$^{-1}$ (143 and 156 meV) appear distinctly only at low temperatures, and are identified as crystal-field $j_{1/2} \to j_{5/2}$ excitations. These correspond to the $\Gamma_6 \to \Gamma_8$ modes, which are the most intense crystal field excitations for $\Delta_2 > 0$ [see Fig.~\ref{fig-theoryraman}(c)]. In principle, there should exist a higher energy $\Gamma_6 \to \Gamma_7$ excitation, but it is predicted to have very small intensity, such that it is unlikely to be resolved in the experimental spectra. It may be noted that an additional broad feature is observed around 1600 cm$^{-1}$ (198 meV). However, the integrated intensity and energy are both too high to correspond to the highest energy $j_{1/2} \to j_{5/2}$ transition.

\begin{figure}[t]
\includegraphics[width=\linewidth]{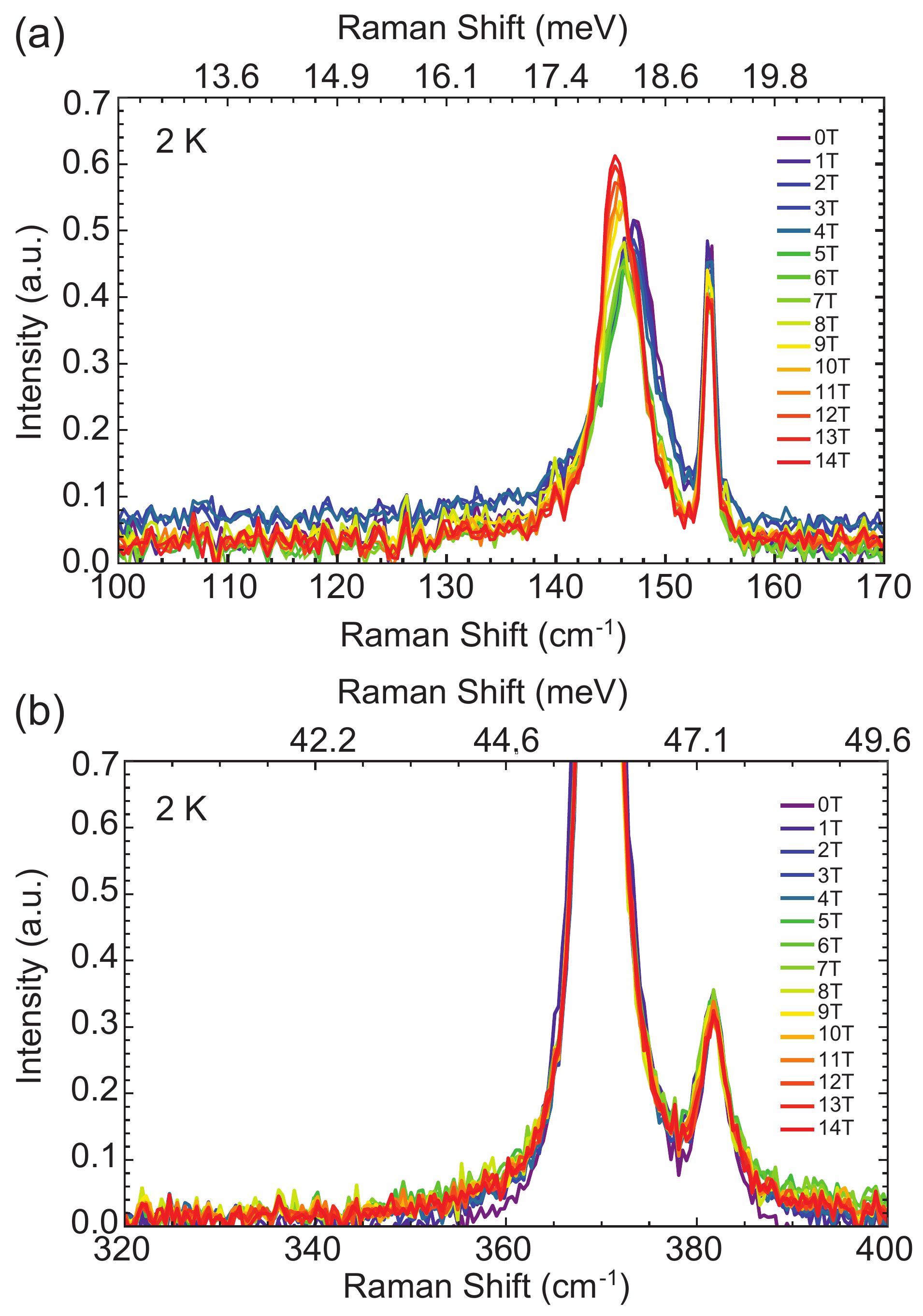}
\caption{Experimental magnetic field-dependent Raman intensities at $T=2$ K for the lowest modes in BCAO in xx polarization configuration, with $B||c$. (a) The shifting of the lower phonon suggests a crystal field excitation in the vicinity of 17.5 meV. (b) The absence of shifting indicates that the modes at 370/380 cm$^{-1}$ are purely phononic.}
\label{fig-bcao-hfield}
\end{figure}

In Fig~\ref{fig-bcao}(b), we show the lower frequency zero-field Raman response. With lowering temperature, we observe the appearance of new peaks at 150 and 380 cm$^{-1}$, indicated by arrows in Fig~\ref{fig-bcao}(b). Fig.~\ref{fig-bcao-hfield} shows magnetic field dependence of these bands at 2~K.  While the application of magnetic H $\parallel$ c does not affect the bands at around 380 cm$^{-1}$, the lower-frequency component of the doublet at 145/150 cm$^{-1}$ shows softening in magnetic field.  This field-dependent frequency shift is unlikely to be the consequence of pure magnetostriction effects, because the other phonons do not shift with field.

The application of magnetic field should lead to a shifting or Zeeman splitting of the crystal-field excitations; the lack of shifting identifies 370/380 cm$^{-1}$ modes as phonons. As noted above, the main phonon at 370 cm$^{-1}$ is essentially a local bending vibration of the [AsO$_4$]$^{3-}$ ions. With respect to the local $C_3$ site symmetry of the arsenate ions, this mode transforms as a doubly degenerate E representation. There are six [AsO$_4$]$^{3-}$ ions per unit cell in the $R\bar{3}$ space group, such that there are overall twelve phonon bands in the solid state associated with this bending vibration, which should be slightly split due to coupling through the Co and Ba layers. It is thus reasonable to expect multiple closely spaced peaks associated with multiple Raman-active arsenate bending modes. In fact, the asymmetry of the main phonon at 370 cm$^{-1}$  at high temperatures suggests that two modes resolved below about 150~K overlap at high temperatures. 

The doublet 145 and 150 cm$^{-1}$ shows a distinct polarization dependence with the lower frequency band following  E$_g$ phonon symmetry, and the higher frequency one, observed only at low temperatures, following  A$_g$ symmetry. While both of these bands were assigned to pure phonon contribution~\cite{zhang2023magnetic}, the sensitivity of the 145~ cm$^{-1}$ mode to magnetic field suggests interpretation of this doublet in terms of vibronic coupling or an overlap  with the the lowest electronic $j_{1/2} \to j_{3/2}$ transition that  falls very near to the 145 cm$^{-1}$ phonon, but has very low natural Raman intensity due to a large $\Delta_2 >0$, see Fig.~\ref{fig-theoryraman}. As discussed below, the existence of the lowest $j_{1/2} \to j_{3/2}$ exciton in this energy range is consistent with the fitting of the higher energy crystal-field transitions, despite the uncertainty in it's precise energy.  There are then two complementary scenarios to explain the shifting of the 145 cm$^{-1}$ phonon: (a) The intensity around 145 cm$^{-1}$ represents a sum of the phonon and weak CF mode; the combined lineshape is modified by the field primarily through shifting of the crystal-field mode. (b) Coupling of the crystal-field excitation to the E$_g$ phonon is symmetry-allowed, resulting in  dynamic vibronic coupling. Vibronic interactions lead to a distinct polarization dependence of the components~\cite{Xu2021} and would explain the lineshape of the 150~cm$^{-1}$ component, which appears temperature independent and lower than all other observed phonons in the spectra of these materials~\cite{zhang2023magnetic}.  The component of the mixed electronic-phonon excitation which carries larger component of electronic wavefunction experiences a shift in magnetic field.

For BCAO, we take a modified fitting approach, employing only the two higher energy $j_{1/2} \to j_{5/2}$ crystal-field excitations whose energies could be precisely measured. We constrain $\lambda = 63$ meV to be consistent with the other materials, and then fit $\Delta_2$. The best fit is obtained for $\Delta_2\text{(fit)} = +40$ meV. This value is consistent with the DFT results. In order to further confirm the validity of the fitted parameters, we also compute the $g$-values. Experimental estimates are $g_\perp\text{(exp)} \sim 5.0$, and $g_{||}\text{(exp)} \sim 2.5$ to 2.7 \cite{regnault2018polarized,halloran2023geometrical}. Our fitted values for $\Delta_2$ and $\lambda$ yield theoretical estimates of $g_\perp = 5.1$ and $g_{||}= 2.4$, which are fully consistent with the experiments, validating the fitted values. On the basis of the fit, we would predict an additional intense electronic transition at $\sim 105$ meV that happens to overlap with the [AsO$_4$]$^{3-}$ stretching vibrations, which explains why it is not separately resolved. We also estimate the lowest $j_{1/2} \to j_{3/2}$ transition to have an energy of $\sim15$ meV, which indeed places it in the correct energy range to contribute to the intensity around, or couple to, the 17.5 meV (145 cm$^{-1}$) phonon. Intensity was also previously observed in this energy range in neutron scattering \cite{halloran2023geometrical}.

While the $j_{1/2}$ ground state doublet moments remains suitable at low energies, the fitted crystal field of $\Delta_2/\lambda \sim 2/3$ is quite large in BCAO. This naturally explains and confirms the assertion of Ref.~\onlinecite{halloran2023geometrical} that the dominant form of exchange anisotropy is a bond-independent easy $ab$-plane anisotropy. It should be emphasized that the lowest $j_{1/2}\to j_{3/2}$ crystal field excitation being at $\sim 15 - 18$ meV places it just above the highest energy one-magnon excitation at the zone edge. Particularly at high field, the magnetic and electronic excitations may therefore overlap. This may be an important consideration for future attempts to refine the magnetic model for BCAO on the basis of fitting neutron scattering data, since level repulsion between the magnetic and electronic modes may significantly alter the magnon dispersions.

{\it Na$_2$Co$_2$TeO$_6$:} In NCTO, the CoO$_6$ octahedra also form an edge-sharing honeycomb lattice. The majority of samples studied (and the ones employed in the present measurements) are found to be in the hexagonal $P6_322$ space group. The magnetic ground state has also been extensively discussed, with leading candidates being zigzag antiferromagnetic order \cite{Lefrancois2016,Bera2017,Samarakoon2021}, and a triple-Q state comprised of a superposition of zigzag wavevectors \cite{yao2023magnetic,chen2021spin,Yao2022}. The material exhibits a complex field-temperature phase diagram \cite{Yao2020,lee2021multistage,xiao2021magnetic,zhang2022electronic,Lin2021} with multiple magnetic and/or structural transitions, showing distinct signatures in e.g.~ESR \cite{Lin2021} and thermal transport \cite{Hong2021,Takeda2022}. However, interpretation of these results is complicated by significant disorder in the interlayer Na positions \cite{Viciu2007,Xiao2019} (which may be partially mitigated in the monoclinic polymorph\cite{dufault2023introducing}). This disorder likely induces a significant randomness in the local crystal fields, which manifests as a rich magnetic excitation spectrum particularly at high fields due to randomness in the $g$-tensors \cite{xiang2023disorder}. Similarly, we expect that disorder may lead to broadening and anomalous lineshapes of the crystal field excitations in NCTO. As such, fitting the spectra produces an approximate model accounting only for the average trigonal distortion.

\begin{figure}[t]
\includegraphics[width=\linewidth]{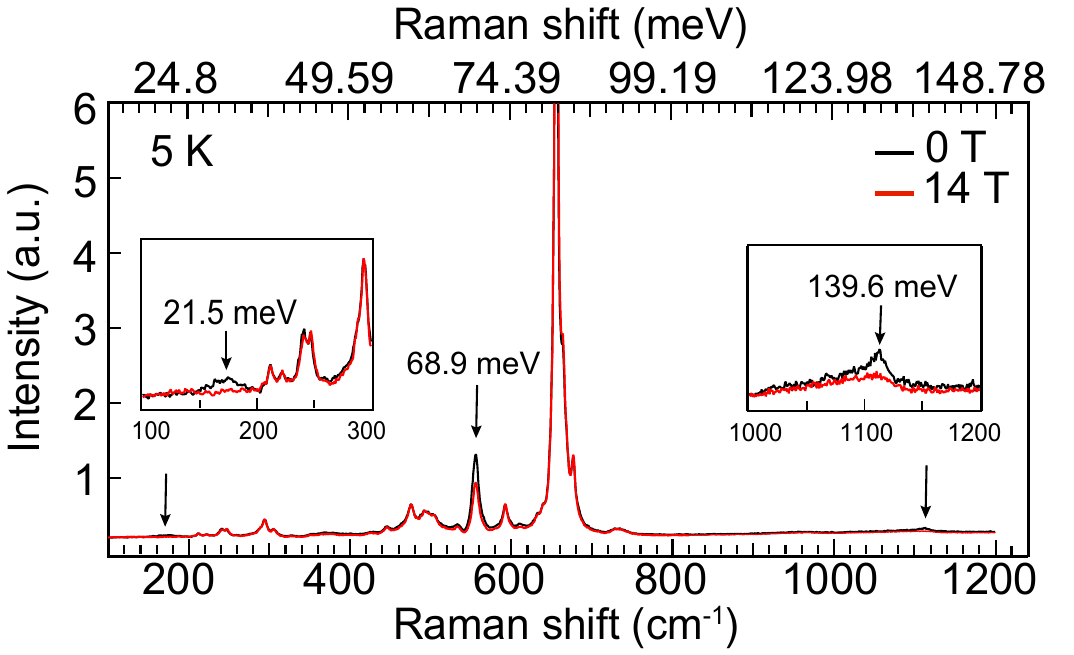}

\caption{Raman spectra measured on Na$_2$Co$_2$TeO$_6$ at 5 K and magnetic field at 0 T (black) and 14 T (red). The black arrows and insets show the Raman peaks that are broadened upon applying magnetic field to 14 T. Data is not polarized.}
\label{fig-ncto}
\end{figure}

Fig.~\ref{fig-ncto} shows low-temperature NCTO Raman spectra. In order to better distinguish possible crystal field excitations, we applied an out-of-plane magnetic field of 14 T, and compare the spectra at zero field. This field is well below the out-of-plane critical field, so that any field-induced changes cannot be associated with magnetostructural phase transitions. With consideration of the random $g$-tensors for the ground and excited states, it is expected that random Zeeman splitting under applied field will lead to a broadening of any mode primarily of electronic origin. Indeed, there are three excitations at 21.5, 68.9, and 137.9 meV with intensity strongly suppressed by the magnetic field -- which distinguishes these modes from the phonons. In addition, a broad band is observed at about 118 meV, as previously reported \cite{chen2021spin}. The lowest energy mode at 21.5 meV  is in agreement with previous infrared measurements \cite{xiang2023disorder} and inelastic neutron scattering \cite{Kim2021}. It is identified as the lower energy $j_{1/2} \to j_{3/2}$ mode.  We identify the high-energy peaks at 118 and 138 meV as $j_{1/2} \to j_{5/2}$ modes. The origin of the intense mode at 69 meV is not completely clear; the strong magnetic field dependence suggests that a crystal field excitation is likely buried under the prominent phonon peak. Indeed, it falls in an expected range for a crystal field excitation; for the purpose of fitting, we tentatively identify 69 meV with the higher energy $j_{1/2} \to j_{3/2}$ mode.    

Fitting these peak energies yields an experimental estimate of $\Delta_2 = +23.7$ meV and $\lambda = 64.4$ meV. In reality, the local Co environments in NCTO are lower symmetry than suggested by a purely trigonal crystal field due to the interlayer Na disorder. The estimated range of off-diagonal contributions to the crystal field from DFT is 2.4 to 18.5 meV, which is somewhat smaller than the fitted value. However, it is clear that the average effective trigonal term is smaller in magnitude for NCTO in comparison with BCAO.

Amongst the honeycomb oxides, we anticipate that the sister compound Na$_3$Co$_2$SbO$_6$ (NCSO) \cite{li2022giant,kang2023honeycomb,vavilova2023magnetic,gu2023easy} may be a more ideal platform than BCAO and NCTO for observing the effects of bond-dependent anisotropic exchange. NCSO is not subject to the Na disorder present in NCTO, and the lowest spin-orbit exciton lies above that of NCTO \cite{Kim2021}, implying a weaker trigonal field. A recent X-ray dichroism study \cite{van2023electronic} constrained $\Delta_2$ for NCSO to be in the range +12 to +20 meV. A future Raman study could provide tighter constraints. 

{\it CoX$_2$:} Finally, for completeness, we place previously studied CoX$_2$ materials (with X = Cl, Br, and I) in context with the oxides studied in this work. The CoX$_2$ family all crystallize in an edge-sharing triangular lattice. Of these, CoCl$_2$ and CoBr$_2$ order ferromagnetically in the 2D planes, with moments oriented preferentially in the planes, suggestive of XY-like anisotropy \cite{magarino1977field}. Early ESR \cite{jacobs1968field} inelastic neutron scattering investigations \cite{hutchings1973neutron,yoshizawa1980neutron} suggested $J_{||}/J_\perp \sim 0.15 - 0.4$ for CoCl$_2$ and $J_{||}/J_\perp \sim 0.4$ CoBr$_2$. CoI$_2$ adopts a spiral magnetic order, also with moments primarily oriented in-plane \cite{kuindersma1981magnetic,kim2023bond}. A recent investigation \cite{kim2023bond} revealed significant bond-dependent anisotropic exchange, in addition to weak global XXZ anisotropy, with $J_{||}/J_\perp \sim 0.95$, suggesting CoI$_2$ to have the weakest trigonal effects. The bond-dependent exchange was particularly implicated to explain strong magnon decay effects. Spectroscopic confirmation of these trends is relevant for solidifying this interpretation.

For the electronic Raman scattering, we consider previously reported data from Ref.~\onlinecite{christie1971electronic,christie1975electronic,mischler1987raman,lockwood1979raman} within our fitting scheme. In these works, the majority of peaks could be clearly resolved, in part due to a sparser phonon spectrum. A similar analysis of the crystal field was performed in these works, but is repeated here for comparison with the oxide materials. On the basis of reported intensities, it is clear that all CoX$_2$ materials have $\Delta_2 >0$. In particular, the experimental fits yield $\Delta_2^{\rm CoCl2} = +18.3$ meV, $\Delta_2^{\rm CoBr2} = +9.9$ meV, and $\Delta_2^{\rm CoI2} = +6.7$ meV. These trends are in full agreement with the trends in reported XY-anisotropies, demonstrating the utility of electronic Raman scattering. On the basis of the fitted values, we also computed the $g$-tensors. For CoCl$_2$, we estimate $g_{||} = 3.0, g_\perp = 5.1$. For CoBr$_2$, we estimate $g_{||} = 3.6, g_\perp = 4.9$. For CoI$_2$, we estimate $g_{||} = 3.8, g_\perp = 4.8$. These are in essential agreement with the limited experimental values: $g_\perp \approx 5.3$ for CoCl$_2$ and $g_\perp \approx 4.91$ for CoBr$_2$. \cite{mischler1987raman}

We note that the experimental $\Delta_2$ values are in stark contrast with the DFT results, which preserve the experimental trend $\Delta_2^{\rm CoCl2} > \Delta_2^{\rm CoBr2} > \Delta_2^{\rm CoI2}$, but find the opposite sign of the trigonal term for all materials. As a consequence, while DFT performs adequately for the oxides, it is not reliable for estimating the crystal-field terms for the halides.

\section{Discussion}\label{sec_conclusions}
In this work, we have made a comprehensive study of the spin-orbital excitons in a series of Co$^{2+}$ materials of interest as quantum magnets, with a view to refine model parameters. Overall, the results emphasize the need to understand the local single-ion effects, which compete strongly with spin-orbit coupling to define the ultimate magnetic couplings between $j_{1/2}$ moments. The main conclusions are as follows:

(i) The trigonal crystal field in BCAO is likely sufficiently strong to place the material on the verge of inapplicability of the $j_{1/2}$ picture. That is, the lowest crystal field excitation likely appears close in energy to the highest magnon. 

(ii) Of all the materials studied, CoI$_2$ represents the most ideal platform for studying bond-dependent anisotropic exchange in Co$^{2+}$ materials. It has the smallest trigonal perturbations, and thus the most ideal $j_{1/2}$ moments. However, we may note, in addition to the well-studied anisotropic terms associated primarily with the exchange between the $t_{2g}$ electrons \cite{Sano2018PRB,Liu2018,Liu2020PRL,winter2022magnetic}, that the strong SOC of the iodine is likely to play a significant role. This generates significant anisotropies for exchange processes involving $e_g$ electrons \cite{stavropoulos2019microscopic,riedl2022microscopic}.

(iii) While the DFT (PBE+Wannier fitting) estimates of the crystal field terms were remarkably accurate for the oxides studied, the experimental values were not accurately reproduced for the Co halide materials. While it may be possible this discrepancy can be fixed in the future by judicious choice of DFT methods, it is highlighted here as a consideration for future {\it ab-initio} studies. The magnetic couplings are sufficiently sensitive to small variations in the single-ion terms that their accurate modelling is a necessity for accurate description of the magnetic model. 

Overall, Raman scattering is a key diagnostic tool for quantum magnets with complex spin-orbital structures. Careful consideration of the crystal-field (spin-orbital) excitons -- their relative Raman intensities and energies -- provides insight into the local single-ion terms, which play a dominant role in several Co$^{2+}$ materials.

\begin{acknowledgments} 
We thank Ramesh Dhakal for insightful discussions. DFT calculations were performed using the Wake Forest University (WFU) High Performance Computing Facility, a centrally managed computational resource available to WFU researchers including faculty, staff, students, and collaborators \cite{WakeHPC}. Work at JHU and the crystal growth at Princeton University were supported as part of the Institute for Quantum Matter, an Energy Frontier Research Center funded by the U.S. Department of Energy, Office of Science, Basic Energy Sciences under Award No.~DE-SC0019331. The NCTO crystal growth at UTK and the sample characterization at GT were both supported by the U.S. Department of Energy (Grant Nos. DE-SC0020254 and DE-SC0023455, respectively). The magneto-Raman measurements supported by the US Department of Energy (DE-FG02-07ER46451) were performed at NHMFL, which is supported by the NSF Cooperative Agreement (Nos. DMR-1644779 and DMR-2128556) and the State of Florida.

\end{acknowledgments}

\bibliographystyle{apsrev}
\bibliography{ref}

\end{document}